\documentstyle[psfig,12pt]{article}
\begin{document}
%%%%%%%%%%%%%%%%%%%%%%%%%%%%%%%%%%%%%%%%%%%%%%%
\newcommand {\ignore}[1]{}
\newcommand{\nota}[1]{\makebox[0pt]{\,\,\,\,\,/}#1}
\newcommand{\notp}[1]{\makebox[0pt]{\,\,\,\,/}#1}
\newcommand{\braket}[1]{\mbox{$<$}#1\mbox{$>$}}
\newcommand{\Frac}[2]{\frac{\displaystyle #1}{\displaystyle #2}}
\renewcommand{\arraystretch}{1.5}
\newcommand{\noi}{\noindent}
\newcommand{\bc}{\begin{center}}
\newcommand{\ec}{\end{center}}
\newcommand{\epm}{e^+e^-}
\def\ne{\hbox{$\nu_e$ }}
\def\nm{\hbox{$\nu_\mu$ }}
\def\bC{\mathop{\bf C}}
\def\eq#1{{eq. (\ref{#1})}}
\def\Eq#1{{Eq. (\ref{#1})}}
\def\Eqs#1#2{{Eqs. (\ref{#1}) and (\ref{#2})}}
\def\Eqs#1#2#3{{Eqs. (\ref{#1}), (\ref{#2}) and (\ref{#3})}}
\def\Eqs#1#2#3#4{{Eqs. (\ref{#1}), (\ref{#2}), (\ref{#3}) and (\ref{#4})}}
\def\eqs#1#2{{eqs. (\ref{#1}) and (\ref{#2})}}
\def\eqs#1#2#3{{eqs. (\ref{#1}), (\ref{#2}) and (\ref{#3})}}
\def\eqs#1#2#3#4{{eqs. (\ref{#1}), (\ref{#2}), (\ref{#3}) and (\ref{#4})}}
\def\fig#1{{Fig. (\ref{#1})}}
\def\lie{\hbox{\it \$}} % fancy L for the Lie derivative
\def\partder#1#2{{\partial #1\over\partial #2}}
\def\secder#1#2#3{{\partial^2 #1\over\partial #2 \partial #3}}
\def\bra#1{\left\langle #1\right|}
\def\ket#1{\left| #1\right\rangle}
\def\VEV#1{\left\langle #1\right\rangle}
\let\vev\VEV
\def\gdot#1{\rlap{$#1$}/}
\def\abs#1{\left| #1\right|}
\def\pri#1{#1^\prime}
\def\ltap{\raisebox{-.4ex}{\rlap{$\sim$}} \raisebox{.4ex}{$<$}}
\def\gtap{\raisebox{-.4ex}{\rlap{$\sim$}} \raisebox{.4ex}{$>$}}
\def\lsim{\raise0.3ex\hbox{$\;<$\kern-0.75em\raise-1.1ex\hbox{$\sim\;$}}}
\def\gsim{\raise0.3ex\hbox{$\;>$\kern-0.75em\raise-1.1ex\hbox{$\sim\;$}}}
\def\contract{\makebox[1.2em][c]{
        \mbox{\rule{.6em}{.01truein}\rule{.01truein}{.6em}}}}
\def\half{{1\over 2}}
\def\bel{\begin{letter}}
\def\eel{\end{letter}}
\def\beq{\begin{equation}}
\def\eeq{\end{equation}}
\def\bef{\begin{figure}}
\def\eef{\end{figure}}
\def\bet{\begin{table}}
\def\eet{\end{table}}
\def\bea{\begin{eqnarray}}
\def\ba{\begin{array}}
\def\ea{\end{array}}
\def\bi{\begin{itemize}}
\def\ei{\end{itemize}}
\def\ben{\begin{enumerate}}
\def\een{\end{enumerate}}
\def\ra{\rightarrow}
\def\ot{\otimes}
\def\lrover#1{
        \raisebox{1.3ex}{\rlap{$\leftrightarrow$}} \raisebox{ 0ex}{$#1$}}
\def\com#1#2{
        \left[#1, #2\right]}
\def\eea{\end{eqnarray}}
\def\bentarrow{\:\raisebox{1.3ex}{\rlap{$\vert$}}\!\rightarrow}
\def\longbent{\:\raisebox{3.5ex}{\rlap{$\vert$}}\raisebox{1.3ex}
        {\rlap{$\vert$}}\!\rightarrow}
\def\onedk#1#2{
        \begin{equation}
        \begin{array}{l}
         #1 \\
         \bentarrow #2
        \end{array}
        \end{equation}
                }
\def\dk#1#2#3{
        \begin{equation}
        \begin{array}{r c l}
        #1 & \rightarrow & #2 \\
         & & \bentarrow #3
        \end{array}
        \end{equation}
                }
\def\dkp#1#2#3#4{
        \begin{equation}
        \begin{array}{r c l}
        #1 & \rightarrow & #2#3 \\
         & & \phantom{\; #2}\bentarrow #4
        \end{array}
        \end{equation}
                }
\def\bothdk#1#2#3#4#5{
        \begin{equation}
        \begin{array}{r c l}
        #1 & \rightarrow & #2#3 \\
         & & \:\raisebox{1.3ex}{\rlap{$\vert$}}\raisebox{-0.5ex}{$\vert$}
        \phantom{#2}\!\bentarrow #4 \\
         & & \bentarrow #5
        \end{array}
        \end{equation}
                }
\def\ap#1#2#3{           {\it Ann. Phys. (NY) }{\bf #1} (19#2) #3}
\def\arnps#1#2#3{        {\it Ann. Rev. Nucl. Part. Sci. }{\bf #1} (19#2) #3}
\def\cnpp#1#2#3{        {\it Comm. Nucl. Part. Phys. }{\bf #1} (19#2) #3}
\def\apj#1#2#3{          {\it Astrophys. J. }{\bf #1} (19#2) #3}
\def\app#1#2#3{          {\it Astropart. Phys. }{\bf #1} (19#2) #3}
\def\asr#1#2#3{          {\it Astrophys. Space Rev. }{\bf #1} (19#2) #3}
\def\ass#1#2#3{          {\it Astrophys. Space Sci. }{\bf #1} (19#2) #3}
\def\aa#1#2#3{          {\it Astron. \& Astrophys.  }{\bf #1} (19#2) #3}
\def\apjl#1#2#3{         {\it Astrophys. J. Lett. }{\bf #1} (19#2) #3}
\def\ap#1#2#3{         {\it Astropart. Phys. }{\bf #1} (19#2) #3}
\def\ass#1#2#3{          {\it Astrophys. Space Sci. }{\bf #1} (19#2) #3}
\def\jel#1#2#3{         {\it Journal Europhys. Lett. }{\bf #1} (19#2) #3}
\def\ib#1#2#3{           {\it ibid. }{\bf #1} (19#2) #3}
\def\nat#1#2#3{          {\it Nature }{\bf #1} (19#2) #3}
\def\nps#1#2#3{          {\it Nucl. Phys. B (Proc. Suppl.) }
                         {\bf #1} (19#2) #3}
\def\np#1#2#3{           {\it Nucl. Phys. }{\bf #1} (19#2) #3}
\def\pl#1#2#3{           {\it Phys. Lett. }{\bf #1} (19#2) #3}
\def\pr#1#2#3{           {\it Phys. Rev. }{\bf #1} (19#2) #3}
\def\prep#1#2#3{         {\it Phys. Rep. }{\bf #1} (19#2) #3}
\def\prl#1#2#3{          {\it Phys. Rev. Lett. }{\bf #1} (19#2) #3}
\def\pw#1#2#3{          {\it Particle World }{\bf #1} (19#2) #3}
\def\ptp#1#2#3{          {\it Prog. Theor. Phys. }{\bf #1} (19#2) #3}
\def\jppnp#1#2#3{         {\it J. Prog. Part. Nucl. Phys. }{\bf #1} (19#2) #3}
\def\rpp#1#2#3{         {\it Rep. on Prog. in Phys. }{\bf #1} (19#2) #3}
\def\ptps#1#2#3{         {\it Prog. Theor. Phys. Suppl. }{\bf #1} (19#2) #3}
\def\rmp#1#2#3{          {\it Rev. Mod. Phys. }{\bf #1} (19#2) #3}
\def\zp#1#2#3{           {\it Zeit. fur Physik }{\bf #1} (19#2) #3}
\def\fp#1#2#3{           {\it Fortschr. Phys. }{\bf #1} (19#2) #3}
\def\Zp#1#2#3{           {\it Z. Physik }{\bf #1} (19#2) #3}
\def\Sci#1#2#3{          {\it Science }{\bf #1} (19#2) #3}
\def\n.c.#1#2#3{         {\it Nuovo Cim. }{\bf #1} (19#2) #3}
\def\r.n.c.#1#2#3{       {\it Riv. del Nuovo Cim. }{\bf #1} (19#2) #3}
\def\sjnp#1#2#3{         {\it Sov. J. Nucl. Phys. }{\bf #1} (19#2) #3}
\def\yf#1#2#3{           {\it Yad. Fiz. }{\bf #1} (19#2) #3}
\def\zetf#1#2#3{         {\it Z. Eksp. Teor. Fiz. }{\bf #1} (19#2) #3}
\def\zetfpr#1#2#3{         {\it Z. Eksp. Teor. Fiz. Pisma. Red. }{\bf #1}
(19#2) #3}
\def\jetp#1#2#3{         {\it JETP }{\bf #1} (19#2) #3}
\def\mpl#1#2#3{          {\it Mod. Phys. Lett. }{\bf #1} (19#2) #3}
\def\ufn#1#2#3{          {\it Usp. Fiz. Naut. }{\bf #1} (19#2) #3}
\def\sp#1#2#3{           {\it Sov. Phys.-Usp.}{\bf #1} (19#2) #3}
\def\ppnp#1#2#3{           {\it Prog. Part. Nucl. Phys. }{\bf #1} (19#2) #3}
\def\cnpp#1#2#3{           {\it Comm. Nucl. Part. Phys. }{\bf #1} (19#2) #3}
\def\ijmp#1#2#3{           {\it Int. J. Mod. Phys. }{\bf #1} (19#2) #3}
\def\ic#1#2#3{           {\it Investigaci\'on y Ciencia }{\bf #1} (19#2) #3}
\def\tp{these proceedings}
\def\pc{private communication}
\def\opc{\hbox{{\sl op. cit.} }}
\def\ip{in preparation}
\relax
%%%%%%%%%%%%%%%%%%%%%%%%%%%%%%%%%%%%
\topmargin -2cm
\textwidth 17cm
\textheight 25.5cm
\evensidemargin  0cm
\def\e{\mbox{e}}
\def\sgn{{\rm sgn}}
\def\gsim{\;
\raise0.3ex\hbox{$>$\kern-0.75em\raise-1.1ex\hbox{$\sim$}}\;
}
\def\lsim{\;
\raise0.3ex\hbox{$<$\kern-0.75em\raise-1.1ex\hbox{$\sim$}}\;
}
\def\MeV{\rm MeV}
\def\eV{\rm eV}
\thispagestyle{empty}
%%%%%%%%%%%%%%%%%%%%%%%%%%%%%%%%%%%%%%%%%%%%%%%%%%%%%%%%%%%%%%%%%%%
%More strong requirements to formatting  than  conventional
%%%%%%%%%%%%%%%%%%%%%%%%%%%%%%%%%%%%%%%%%%%%%%%%%%%%%%%%%%%%%%%%%%%
\topmargin -27pt
\textwidth 6in
\textheight 24.5cm
\renewcommand{\baselinestretch}{1.25}
\oddsidemargin 5mm
\parfillskip0pt plus\textwidth\relax
\widowpenalty500
\clubpenalty500
%%%%%%%%%%%%%%%%%%%%%%%%%%%%%%%%%%%%%%%%%%%%%%%%%%%%%%%%%%%%%%%%%%%
\makeatletter
%%%%%%%%%%%%%%%%%%%%%%%%%%%%%%%%%%%%%%%%%%%%%%%%%%%%%%%%%%%%%%%%%%%
\newdimen\normalarrayskip              % skip between lines
\newdimen\minarrayskip                 % minimal skip between lines
\normalarrayskip\baselineskip
\minarrayskip\jot
\newif\ifold             \oldfalse
\newif\ifdisplayarray    \displayarraytrue
\newif\ifbigarray        \bigarraytrue
\def\arraymode{\ifold\relax\else\ifdisplayarray\displaystyle\else\relax\fi\fi}
% mode of array enrties
\def\eqnumphantom{\phantom{(\theequation)}}     % right phantom in eqnarray
\def\@arrayskip{\ifold\baselineskip\z@\lineskip\z@\else\ifbigarray
     \baselineskip\normalarrayskip\lineskip\minarrayskip
     \else
     \baselineskip\z@\lineskip\z@\fi\fi}
\def\@arrayclassz{\ifcase \@lastchclass \@acolampacol \or
\@ampacol \or \or \or \@addamp \or
   \@acolampacol \or \@firstampfalse \@acol \fi
\edef\@preamble{\@preamble
  \ifcase \@chnum
     \hfil$\relax\arraymode\@sharp$\hfil
     \or $\relax\arraymode\@sharp$\hfil
     \or \hfil$\relax\arraymode\@sharp$\fi}}
\def\@array[#1]#2{\setbox\@arstrutbox=\hbox{\vrule
     height\arraystretch \ht\strutbox
     depth\arraystretch \dp\strutbox
     width\z@}\@mkpream{#2}\edef\@preamble{\halign \noexpand\@halignto
\bgroup \tabskip\z@ \@arstrut \@preamble \tabskip\z@ \cr}%
\let\@startpbox\@@startpbox \let\@endpbox\@@endpbox
  \if #1t\vtop \else \if#1b\vbox \else \vcenter \fi\fi
  \bgroup \let\par\relax
  \let\@sharp##\let\protect\relax
  \@arrayskip\@preamble}
%
%
%  standard \eqnarray, but middle element in \displaystyle
%
%
\def\eqnarray{\stepcounter{equation}
              \let\@currentlabel=\theequation
              \global\@eqnswtrue
              \global\@eqcnt\z@
              \tabskip\@centering
              \let\\=\@eqncr
              $$%
 \halign to \displaywidth\bgroup
    \eqnumphantom\@eqnsel\hskip\@centering
    $\displaystyle \tabskip\z@ {##}$%
    &\global\@eqcnt\@ne \hskip 2\arraycolsep
         \hfil$\arraymode{##}$\hfil
    &\global\@eqcnt\tw@ \hskip 2\arraycolsep
         $\displaystyle\tabskip\z@{##}$\hfil
         \tabskip\@centering
    &{##}\tabskip\z@\cr}
%
%
%   \marray     using:  \begin{marray}{rcl} ... \end{marray}
%
\newenvironment{marray}{\begin{equation}\begin{array}}%
{\end{array}\end{equation}}
%
%   \carray     using:  \begin{carray} ... \end{carray}
%
\newenvironment{carray}{\begin{equation}\begin{array}{rcl}}%
{\end{array}\end{equation}}
\def\be{\@ifnextchar[{\def\ee{\end{equation}}\begin{equation}\l@b}%
{\def\ee{$$}$$}}
\def\l@b[#1]{\label{#1}}
\def\ba{\@ifnextchar[{\def\ee{\end{carray}}\begin{carray}\l@b}%
{\def\ee{\end{array}$$}$$\begin{array}{rcl}}}
\def\barray#1{\@ifnextchar[{\def\ee{\end{marray}}\begin{marray}{#1}\l@b}%
{\def\ee{\end{array}$$}$$\begin{array}{#1}}}
%
%     \herring{....}
%        - analog of \tabbing environment, but for math mode
%        - It's strange, that LaTeX haven't anything similar.
%       Why herring?  I don't know.
%
%
\def\herring{\@ifnextchar[{\@herring}{\@herring[\vcenter]}}
\def\@herring[#1]#2{\begingroup
\def\*{\\ \>}
\topsep0pt
\partopsep0pt
\def\tabbing{\lineskip\jot \lineskiplimit\jot
     \let\>\@rtab\let\<\@ltab\let\=\@settab
     \let\+\@tabplus\let\-\@tabminus\let\`\@tabrj\let\'\@tablab
     \let\\=\@tabcr
     \global\@hightab\@firsttab
     \global\@nxttabmar\@firsttab
     \dimen\@firsttab\@totalleftmargin
     \global\@tabpush0 \global\@rjfieldfalse
     \trivlist \item[]\if@minipage\else\vskip\parskip\fi
     \setbox\@tabfbox\hbox{\rlap{\indent\hskip\@totalleftmargin
       \the\everypar}}\def\@itemfudge{\box\@tabfbox}\@startline\ignorespaces}
\def\@startfield{\global\setbox\@curfield\hbox
                    \bgroup$\displaystyle}%
\def\@stopfield{$\egroup}%
#1{\begin{tabbing}#2\end{tabbing}}\endgroup}
%%%%%%%%%%%%%%%%%%%%%%%%%%%%%%%%%%%%%%%%%%%%%%%%%%%%%%%%%%%%%%%%%%%%%%
\def\boldbox#1{{\mathsurround0pt\mathchoice{\hbox{\boldmath $#1$}}
{\hbox{\boldmath $#1$}}{\hbox{\boldmath $\scriptstyle#1$}}
{\hbox{\boldmath $\scriptscriptstyle#1$}}}}
\def\textbox#1{{\mathchoice{\mbox{#1}}{\mbox{#1}}
{\mbox{{\scriptsize#1}}}{\mbox{{\tiny#1}}}}}
\def\sect#1{\ref{#1}}
\def\eq#1{(\ref{#1})}
\def\theequation{\thesection.\arabic{equation}}
\@addtoreset{equation}{section}
\def\@cite#1#2{\hbox{ [#1\if@tempswa ,#2\fi]}}
\def\@citex[#1]#2{\if@filesw\immediate\write\@auxout{\string\citation{#2}}\fi
  \def\@citea{}\@cite{\@for\@citeb:=#2\do
    {\@citea\def\@citea{,\penalty\@m}\@ifundefined  % space removed
       {b@\@citeb}{{\bf ?}\@warning
       {Citation `\@citeb' on page \thepage \space undefined}}%
\hbox{\csname b@\@citeb\endcsname}}}{#1}}
%%%%%%%%%%%%%%%%%%%%%%%%%%%%%%%%%%%%%%%%%%%%%%%%%%%%%%%%%%%%%%%%%%%%%
\def\@sect#1#2#3#4#5#6[#7]#8{\ifnum #2>\c@secnumdepth
     \def\@svsec{}\else
     \refstepcounter{#1}\edef\@svsec{\csname the#1\endcsname.%
     \hskip 0.8em }\fi
     \@tempskipa #5\relax
      \ifdim \@tempskipa>\z@
        \begingroup #6\relax
          \@hangfrom{\hskip #3\relax\@svsec}{\interlinepenalty \@M #8\par}%
        \endgroup
       \csname #1mark\endcsname{#7}\addcontentsline
         {toc}{#1}{\ifnum #2>\c@secnumdepth \else
                      \protect\numberline{\csname the#1\endcsname}\fi
                    #7}\else
        \def\@svsechd{#6\hskip #3\@svsec #8\csname #1mark\endcsname
                      {#7}\addcontentsline
                           {toc}{#1}{\ifnum #2>\c@secnumdepth \else
                             \protect\numberline{\csname the#1\endcsname}\fi
                       #7}}\fi
     \@xsect{#5}}
\newenvironment{appendices}{\begingroup
\setcounter{subsection}{0}
\def\thesubsection{\Alph{subsection}}%
\def\theequation{\thesubsection.\arabic{equation}}%
\@addtoreset{equation}{subsection}%
\def\appendix##1{\subsection{##1}}%
}{\endgroup}
\makeatother
%%%%%%%%%%%%%%%%%%%%%%%%%%%%%%%%%%%%%%%%%%%%%%%%%%%%%%%%%%%%%%%%%%%%%%
\def\note#1{\typeout{#1}}
\def\BUG#1{\vrule width 2pt height 8pt depth 2pt\relax
\typeout{BUG? page= \thepage: #1}}
\def\dubious{\typeout{DUBIOUS: page= \thepage}}
%%%%%%%%%%%%%%%%%%%%%%%%%%%%%%%%%%%%%%%%%%%%%%%%%%%%%%%%%%%%%%%%
\topmargin -2cm
\textwidth 13.5cm
\textheight 25.5cm
\evensidemargin  0cm
\def\e{\mbox{e}}
\def\sgn{{\rm sgn}}
\def\gsim{\;
\raise0.3ex\hbox{$>$\kern-0.75em\raise-1.1ex\hbox{$\sim$}}\;
}
\def\lsim{\;
\raise0.3ex\hbox{$<$\kern-0.75em\raise-1.1ex\hbox{$\sim$}}\;
}
\def\MeV{\rm MeV}
\def\eV{\rm eV}
%\tableofcontents
\thispagestyle{empty}
\begin{titlepage}
\today
\begin{center}
\hfill hep-ph\\
\hfill FTUV/95-41\\
\hfill IFIC/95-43\\
\vskip 0.3cm
\LARGE
{\bf A New Type of Resonant Neutrino Conversions Induced by Magnetic Fields}
\end{center}
\normalsize
\vskip1cm
\begin{center}
{\bf S. Sahu},
{\bf V.B.Semikoz}
\footnote{On leave from the {\sl Institute of the
Terrestrial Magnetism, the Ionosphere and Radio
Wave Propagation of the Russian Academy of Sciences,
IZMIRAN, Troitsk, Moscow region, 142092, Russia}.}
{\bf and J. W. F. Valle}
\footnote{E-mail valle@flamenco.ific.uv.es}\\
\end{center}
\begin{center}
\baselineskip=13pt
{\it Instituto de F\'{\i}sica Corpuscular - C.S.I.C.\\
Departament de F\'{\i}sica Te\`orica, Universitat de Val\`encia\\}
\baselineskip=12pt
{\it 46100 Burjassot, Val\`encia, SPAIN         }\\
\vglue 0.8cm
\end{center}

\begin{abstract}

We consider resonant neutrino conversions
in magnetised matter, such as a degenerate electron gas.
We show how magnetisation effects caused by axial vector
interactions of neutrinos with the charged leptons in the
medium can induce a new type of resonant neutrino conversion
which may occur even in situations where the MSW effect
does not occur, such as the case of degenerate or inverted
neutrino mass spectra. Our new resonance may simultaneously
affect anti-neutrino $\bar{\nu_a} \leftrightarrow \bar{\nu}_b$
as well as neutrino $\nu_{a} \leftrightarrow \nu_b$ flavour
conversions, and therefore it may substantially affect
supernova neutrino energy spectra.
Using SN1987A data we conclude that only laboratory
experiments with long baseline such as ICARUS or MINOS
are likely to find neutrino oscillations due to their
sensitivity to small $\Delta m^2$. We also comment on
the possibility of resonant conversions induced by
Majorana neutrino transition moments and mention the
case of sterile neutrinos $\nu_s$.

\end{abstract}
\vfill
\end{titlepage}
\newpage

\section{Introduction}

Neutrino propagation in media with random magnetic fields
has attracted considerable attention recently, both from
the point of view of the early universe cosmology
as well as astrophysics \cite{Zeldo}.
The presence of random magnetic fields in cosmology
as well as in various astrophysical objects can strongly
affect neutrino conversion rates and this could have
important implications.

Some recent papers have considered neutrino propagation
in media with time-varying magnetic field $B(t)$, when the
magnetic field is a regular changing field like a twist
(circularly polarised) one \cite{AnezirisSchechter,Smirnov},
or the linear polarised Alfven wave \cite{Semikoz}. In any
of these regular magnetic field cases neutrino conversions
occur in oscillating regime. In contrast in the case of a
random magnetic field $\langle {\tilde B}(t)\rangle = 0$ with
$\langle {\tilde B}(t)^2\rangle \neq 0$, the neutrino
conversions become aperiodic \cite{SemikozValle}.

In general the magnetic field can be separated into
two parts, $B_j (t) = B_{j0} +{\tilde B}_j(t)$,
the large-scale constant field $B_{j0}$ and a
random field ${\tilde B}_{j}(t)$. One can consider
neutrino propagation in a medium, such as a supernova,
in two simple regimes: $\tilde{B}(t) \gg B_0$ and
$\tilde{B}(t) \ll B_0$. The first case has been
treated in previous works, where the effect of a
strong random magnetic field,
$\sqrt{\langle B^2(t)\rangle} \gg B_{0}$ on
active-sterile supernova neutrino conversions
was discussed \cite{sergio}. In addition,
the corresponding effect of strong random magnetic
field upon neutrino transitions induced by a
transition magnetic moment in the early
universe hot plasma or in a supernova
was discussed in ref. \cite{sergio1}.

In this paper we study a new class of resonant neutrino
conversions in magnetised matter, such as a degenerate
electron gas in the case of strong large-scale
magnetic field.

We demonstrate how the magnetisation effects
caused by axial vector interactions of neutrinos with charged
leptons in the medium can induce a new type of resonant neutrino
conversion which may occur even in kinematical situations where
the MSW effect is forbidden. We discuss as an example the case
of degenerate neutrinos and the interesting case of neutrinos
with a mass difference $\Delta m^2 \sim 10^{-5} \: eV^2$
relevant to the solar neutrino problem. If such neutrinos
undergo MSW conversions in the sun then their supernovae
anti-neutrinos certainly do not. Nevertheless our new
mechanism allows both
$\bar{\nu_a} \leftrightarrow \bar{\nu}_b$
{\sl supernova anti-neutrino} conversions as well as
$\nu_{a} \leftrightarrow \nu_b$
{\sl solar neutrino} conversions to be resonantly enhanced.
Similar effects may exist also for
conversions involving sterile neutrinos $\nu_s$.
The effect of our new resonance in supernovae
anti-neutrino energy spectra leads to constraints
(from SN1987A) on neutrino conversion parameters
$\Delta m^2$ and $\sin^2 2 \theta$. In particular
we show that only $\Delta m^2 \lsim 10^{-3} \: eV^2$
are possible, thus ruling out the possibility of
finding neutrino oscillations at laboratory experiments
with short baseline. Thus, if our assumptions hold
and our resonance takes place in supernovae, then
only the new generation of long baseline accelerator
experiments such as ICARUS and MINOS or reactor experiments
such as CHOOZ or San Onofre are likely to see any effect.

In case of the resonant $\bar{\nu}_{\mu} \leftrightarrow \bar{\nu}_e$
conversions which are suppressed in absence of magnetic field this
allows us to obtain some bounds on neutrino mixing parameters from
the non-observation of distortions of the SN1987A electron
anti-neutrino spectrum in the Kamiokande and IMB experiments.
We also comment on the possibility of such resonant
conversions induced by Majorana neutrino transition moments.

%%%%%%%%%%%%%%%%%%%%%%%%%%%%%%%%%%%%%
\section{Neutrino conversions in magnetised electron gas}
%%%%%%%%%%%%%%%%%%%%%%%%%%%%%%%%%%%%%
\vskip 0.5cm

The Schr\"{o}dinger evolution equation describing
propagation of a system of two Majorana neutrino
species in a magnetic field is given by a four
dimensional evolution Hamiltonian \cite{BFD}.
Here we are interested in the situation where
the neutrinos propagate in a medium. However,
in contrast to ref. \cite{MSW,LAM} we will
include the magnetisation effects due to the
mean axial vector neutrino interactions
with the charged leptons in the medium.
For definiteness we assume two neutrino
species $\nu_a$ where $a = e,\mu,\tau$ denotes a
definite neutrino flavour and $\nu_x$ where  $x=s,b$
denotes either a sterile or an  active  neutrino.
In the following we will specialise our attention to two
simple cases
\ben
\item
neutrino transition magnetic moment is neglected,
$\mu = \mu^{\nu}_{ax}=0$. This is the generalisation of MSW
theory \cite{MSW} in the presence of magnetic field but neglecting
neutrino magnetic moment
\item
neutrino mixing is neglected, $\theta_{ax} \equiv \theta  = 0$.
This is a generalisation of the spin-flavour conversion theory
\cite{LAM} in the presence of non-vanishing mean axial vector
neutrino interactions
\een
In these cases the evolution Hamiltonian simplifies to a two
dimensional evolution one, which may be generically written as
\barray{ll}[initial]
i\frac{d}{dt}\left (\matrix{\nu_a\\\nu_x}\right )
 = \left (\matrix{H_{aa} & H_{ax}\\ H_{xa} & H_{xx} }
          \right )\left (\matrix{\nu_a\\\nu_x}\right ),   ~
\ee
By convention we always choose $H_{xx}=0$.

In case 1 \eq{initial} the  diagonal component $H_{aa}$ is
given by
\be[haa1]
H_{aa}=V_a-\Delta \cos 2\theta + f(\mu_{eff})B_{\parallel}(t)
\ee
while in case 2 it will take the form
\be[haa]
H_{aa}=V_a-\Delta + f(\mu_{eff})B_{\parallel}(t)
\ee
where $\Delta=(m_2^2-m_1^2)/2E$ vanishes for degenerate
Majorana neutrinos and $V_a$ is the appropriate difference
of vector potentials describing the interactions with matter.
Following ref. \cite{SemikozValle} we include the effect of
the mean axial vector current $V^{(axial)}$ of charged
leptons in an external magnetic field $B_{\parallel}$ = {\bf Bq}$/q$,
denoting the relevant term by $ f(\mu_{eff}) B_{\parallel}(t)$.
This term involves the difference of axial vector potentials
of the two neutrinos. Both terms depend on the channel of
neutrino conversions considered. For case 1, we will
consider helicity conserving
$\nu_e \leftrightarrow \nu_{\mu}$ and $\nu_e\leftrightarrow \nu_s$
conversions. For case 2, we will consider helicity flipping
conversions of active neutrinos to both active and sterile neutrinos.

Now we turn to the off-diagonal entry $H_{ax}=H_{xa}$.
For case 1 it takes the form
\be
H_{ax}=\Delta \sin 2\theta/2
\ee
The second case corresponds to the spin-flavour
transitions when we neglect mixing $s \ra 0$ so that
$\cos 2\theta = 1$. In this case the off-diagonal entry becomes
\be
H_{ax} = \mu B_{\perp}(t),
\ee
where $\bf B_{\perp}$ is the component of the magnetic field
transverse to the neutrino momentum {\bf q},
$\mu B_{\perp}(t) \equiv H_{\perp}(t)$.
For Dirac neutrinos $\Delta = 0$ and $\mu$ corresponds to
the corresponding magnetic moment
\footnote{From the two-component point of view the
Dirac neutrino magnetic moment is a particular case
of transition moment, involving an active to a sterile
neutrino.}.

In both cases $V_a$ is the difference of the active neutrino
vector interaction potentials, $V_a = V_{\nu_a} - V_{\nu_x}$,
(for $\nu_a$, $a\neq b$; $a =e, \mu, \tau$). For the sterile
neutrino case $V_{\nu_x} = 0$. For a left-handed electron
neutrino in a supernova with core density
$\rho = \rho_{14}\times 10^{14}$ $g/cm^3$ the vector
potential is
\be[potential2]
V_{\nu_e}\simeq 3.8\times 10^{-6}\rho_{14}f^{(e)}(Y_e)MeV,
\ee
where $f^{(e)}(Y_e) = 3Y_e-1$ is a function of the abundances
for our probe electron neutrino. Here $Y_e=n_e/n_B$ is the
electron abundance, $n_B$ is the baryon density, and we
neglect neutrino background contribution, $Y_{\nu}= 0$.
For the electron anti-neutrino case one would
have $V_{\bar{\nu}_e} = - V_{\nu_e}$.

For a left-handed muon neutrino the potential is
\be[potential3]
V_{\nu_{\mu}}\simeq 3.8\times 10^{-6}\rho_{14}f^{(\mu)}(Y_e)MeV,
\ee
with abundance function $f^{(\mu)}(Y) =Y_e - 1$.
Similarly for the muon anti-neutrino the potential
$V_{{\bar \nu}_{\mu}}=-V_{\nu_{\mu}}$.

Using these neutrinos and anti-neutrino potentials
we can easily calculate the resulting vector potential
($V_a$) for different flavour and (or) spin-flavour
conversions. For $\nu_e \leftrightarrow \nu_{\mu}$
flavour conversion the potential is
\be[potentialMSW]
V_a\simeq 3.8 \times 10^{-6} \rho_{14} 2Y_e MeV
\ee
and for spin-flavour conversions $\bar{\nu}_e\leftrightarrow \nu_{\mu}$,
the potential is
\be[potentialLAM]
V_a\simeq 3.8 \times 10^{-6}\rho_{14} 2 (1 - 2Y_e) MeV.
\ee
For the active-sterile neutrino conversions
$\nu_e\leftrightarrow \bar{\nu_s}$ one gets
\be[potentialas]
V_a\simeq 3.8\times 10^{-6}\rho_{14}(3Y_e - 1)MeV.
\ee

The averaged matrix element for the axial current
denoted by the symbol $<....>_0$ is given by \cite{SemikozValle}
\barray{ll}[axial1]
V^{(axial)}_{\nu_b}
={G_F\over {\sqrt 2}}
\sum_{a=e,\mu,\tau} (-2c^{ba}_A){<{\bar\psi_a}\gamma_z\gamma_5\psi_a>_0}
\\
={ \frac{G_F}{\sqrt 2}}\sum_{a=e,\mu,\tau} (-2c^{ba}_A)
{\mu_B}{2e{\bf B_\parallel}\over (2\pi)^2}\times
\\
\int^{\infty}_0 dp_z
\Big [
{1\over {exp((\sqrt{p_z^2+m_a^2}-\zeta)/T)+1}}
+
{ 1 \over {exp((\sqrt{p_z^2+m_a^2}+\zeta)/T)+1}} \Big ],
\ee
where $\mu_B = e/2m_a$ is the Bohr magneton and $c^{ba}_A =
\mp 0.5$
\footnote{$c_A$ changes sign for corresponding anti-neutrinos. }
is the axial coupling constant (upper sign for $b =a$
and lower one for $b \neq a$). Comparing \eq{axial1} with
$\mu_{eff}B_{\parallel}$ and neglecting the contribution
of positrons, muons and taus in a supernova degenerate
electron gas we can define
\be[momentsupernova]
\mu_{eff} = - 9c_A\times 10^{-13}\mu_B(p_{F_e}/MeV).
\ee
Note that the quantity $\mu_{eff}$ has no relation to a real
magnetic moment since it does not lead to a change of helicity.

In the diagonal entry of $H_{aa}$ the function $f(\mu_{eff})$
is the difference in $\mu_{eff}$ for the two species of
neutrinos considered, which is determined by the value of $c_A$.

For the case of $\nu_e \leftrightarrow \nu_{\mu}$ flavour
conversions the axial contribution doubles in the difference
of $2\mu_{eff}$, so that we obtain
\be[diagonal1]
H_{aa}  = 3.8\times 10^{-6}\rho_{14}2Y_e - \Delta \cos 2\theta +
2\mu_{eff} B_{\parallel}.
\ee
In contrast, for spin-flavour transitions
($\bar{\nu}_e\leftrightarrow \nu_{\mu}$)
the axial contribution cancels in the difference of $\mu_{eff}$,
so that the entry $H_{aa}$ is given by the corresponding neutrino
vector potential \eq{potentialLAM},
\be[diagonal2]
H_{aa} =
3.8\times 10^{-6}\rho_{14} 2 (1 - 2Y_e ) MeV - \Delta.
\ee
As a result of this cancellation the effects of magnetisation
are absent for spin-flip flavour conversions, and will
not consider in what follows.
Finally for the active-sterile neutrino
conversions $\nu_e \leftrightarrow \nu_s$
we obtain for the upper diagonal entry
\be[diagonal3]
H_{aa}  =
3.8\times 10^{-6}\rho_{14}(3Y_e -1) - \Delta \cos 2\theta+
\mu_{eff}B_{\parallel},
\ee
for $s\neq 0$, $\mu = 0$ ($H_{ax} = \Delta \sin 2\theta/2$)
and
\be[diagonal4]
H_{aa}  = 3.8\times 10^{-6}\rho_{14}(3Y_e -1) - \Delta +
\mu_{eff}B_{\parallel},
\ee
for $s = 0$, $\mu \neq 0$ ($H_{ax} = \mu B_{\perp}$).

With the help of the equations derived in this section
many new types of neutrino conversions can be described.
In the next session we will focus in the case of resonant
neutrino flavour conversions in supernovae, neglecting
neutrino transition moments.

%%%%%%%%%%%%%%%%%%%%%%%%%%%%%%%%%%%%%%%%%%%%
\section{
Resonant neutrino conversions in magnetised electron
gas with constant magnetic field}
%%%%%%%%%%%%%%%%%%%%%%%%%%%%%%%%%%%%%%%%%%%%

Consider a strong large-scale magnetic field $B_{0j}(t)$
and let us neglect random components ${\tilde B}_j(t)$,
i.e. $B_{0j}(t) \gg {\tilde B}_j(t)$. Expressing the
magnetic field in terms of its transverse and longitudinal
components we have
\be[magfield]
B^2(t) = B^2(t)\cos^2\alpha +B^2(t)\sin^2\alpha
=B^2_{\parallel}(t)+B^2_{\perp}(t),
\ee
where $\alpha$ is the angle between neutrino momentum
{\bf q} and magnetic field {\bf B}. In this case one can
write the differential equation describing the evolution
of the neutrino flavour conversion probability $\cal{P}$ as
\be[master1]
\ddot{\cal{P}} + \omega_0^2 \cal{P}
= {{\Delta^2 \sin^22\theta}\over 2} \:.
\ee
In \eq{master1} the square of the conversion frequency is
\be[frequency1]
\omega_0^2 = (V - \Delta \cos 2\theta + f(\mu_{eff})B_{\parallel})^2
+ \Delta^2 \sin^22 \theta .
\ee
where $f(\mu_{eff})= 2\mu_{eff}$ for $\nu_e\leftrightarrow \nu_{\mu}$
transitions, $f(\mu_{eff})= 0$ for $\bar{\nu}_e\leftrightarrow \nu_{\mu}$
transitions  and $f(\mu_{eff}) = \mu_{eff}$ for
$\nu_e\leftrightarrow \nu_s$ conversions.

Let us consider a supernova with a strong constant
magnetic field $B$ (${\tilde B_j}(t)  = 0$). In this
case the solution of \eq{master1} for the case of
constant density reduces to
\be[solution4]
\cal{P}(t) =
\frac{\Delta^2 \sin^22 \theta}
{(V - \Delta \cos 2\theta + f(\mu_{eff})B_{\parallel})^2
+ ({\Delta \sin 2\theta})^2}\sin^2(\frac{\omega_0t}{2})
\ee
where the frequency $\omega_0$ is given \eq{frequency1}.

This has clearly a resonant form. The most interesting
case from the point of view of observation is the case
of anti-neutrino flavour conversions. The corresponding
resonant condition can be written as
\be[res1]
V + \Delta \cos 2\theta  + 2 \mu_{eff}B \cos \alpha = 0 \:.
\ee
The above resonance condition can be fulfilled for the
outer layers of a supernova, where the density reduces
to $\rho \simeq 10^5$ $gcm^{-3}$ and the Fermi momentum
$p_{F_e}/MeV \simeq (\rho Y_e/10^7gcm^{-3})^{1/3}$.

Note that this condition can always be fulfilled for some point
along the neutrino trajectory, since the last term can take on
any values for $-1 < \cos \alpha < 1$
\footnote{Note that for isotropic emission, neutrinos always
cross magnetic field force lines. As a result it seems unlikely
to expect $B_{\parallel} = 0$, i.e. that neutrinos will propagate
strictly perpendicular to the magnetic field.}.

It is well-known that anti-neutrino flavour transitions
are suppressed in the absence of a magnetic field for
$\Delta = (m^2_2 - m_1^2)/2E > 0$. In contrast the
third term in \eq{res1} allows us to obtain a resonance.

Note also that our new resonance condition can be fulfilled
even for the case of degenerate neutrinos $m_1 = m_2$ when
$\Delta = 0$ and the MSW resonance \cite{MSW} is absent.
This is similar to the mechanism described in ref. \cite{massless}.
For this $\Delta = 0$ case \eq{res1} simplifies to
\be[resonance]
\Bigl (\frac {|B\cos\alpha|}{10^{12}G}\Bigr )^{3/2} \simeq
17 Y_e \rho_5.
\ee
The role of the non-universal interaction in ref.
\cite{massless} is played here by the axial vector
interactions of neutrinos with the charged leptons
of the medium. If \eq{resonance} is fulfilled,
analogously to ref. \cite{massless}, the resonance
will take place simultaneously for both anti-neutrino
as well as neutrino channels.

Now we discuss the issue of adiabaticity of our
resonant conversion. We can define an adiabaticity
parameter at resonance as follows:
\be[adiabaticity1]
\kappa =  \frac{2(\Delta \sin 2\theta)^2}{\mid dV/dr +
2 d(\mu_{eff}B_{\parallel})/dr\mid },
\ee
for $\nu_e \leftrightarrow \nu_{\mu}$ and
$\bar{\nu}_e \leftrightarrow \bar{\nu}_{\mu}$.
Adiabatic neutrino conversions would require
the adiabaticity parameter \eq{adiabaticity1} to be
very large at resonance, $\kappa_R \gg 1$.
As we show below this condition is not
fullfiled in our case. Indeed, note that
for a fixed electron abundance $Y_e \approx constant$,
the profile of the electron density is the same as
the matter density profile, which is assumed to be
$\rho\sim r^{-3}$, so that $p_{F_e} \sim r^{-1}$.
Thus for negligible $\Delta \simeq 0$ and at resonance
the adiabaticity parameter \eq{adiabaticity1}
becomes infinity $\kappa_R = \infty$ if we have
conservation of the magnetic field flux
\footnote{From Gauss theorem $div${\bf B}$= 0$,
we expect the magnetic field profile to be
$B \simeq 10^{12} G/(10km/r)^2$.}, so that $B\sim r^{-2}$
and $\rho \sim r^{-3}$. This would hold irrespective
of the magnetic field and density profiles.
Similarly, this will be the case for dipole type regular
magnetic field $B \sim r^{-3}$ if $\rho \sim r^{-4.5}$.

In reality, during the first seconds of the main neutrino
burst the density drops more steeply \cite{Turner}
\be[densityprofile]
\rho = \rho_0 \Bigl (\frac{30km}{r}\Bigr )^5 \:.
\ee
where $\rho_0 = 10^{12} g/cm^3$. We may describe the
magnetic field profile via the scaling index $m$ defined as
\be[magprofile]
B = B_0\Bigl (\frac{10km}{r}\Bigr )^m \:.
\ee
Let us note that the magnetic field profile \eq{magprofile}
is more speculative than of \eq{densityprofile}. For example
there has been so far no clear X-ray observations for SN1987A.
Below we assume different scaling indexes $m = 1/2, 1, 3/2$
for the {\it mean random magnetic field} corresponding to
different large scale ($r \sim R_{res}\sim 700 $ $km$)
field structure. For instance, $m=3/2$ is appropriate
for 3-d domains like dipole magnetic fields with size
of order $L_0\sim 1$ km randomly oriented in the
supernova \cite{DunkanThomson}
\footnote{Note that the large-scale form \eq{magprofile}
of the instantaneous {\it mean} field consisting of such
dipoles has a completely different profile than the individual
dipoles. As shown in ref. \cite{Hogan} it is characterised by
the scaling index $m = 3/2$ }.

Unfortunately this profile is compatible with our
resonance condition \eq{resonance} only for very large field
values at the supernova core, $B_0\sim 3\times 10^{16}$ $G$
\cite{DunkanThomson}.

On the other hand the case $m= 1/2$ corresponds to
$B_0\sim 3\times 10^{13}$ Gauss at the core, which is
acceptable for a magnetised neutron star. While we
lack a compelling physical motivation for such
an exotic random magnetic field profile, it is
not in conflict with any observational fact.
It would correspond to filaments like super-conducting
needles aligned along the neutrino trajectory
\footnote{D. D. Sokoloff, private communication}.

As we will show below, we find that our neutrino
parameter bounds are not sensitive to the magnetic
field structure \eq{magprofile} which is crucial
for resonant condition \eq{resonance} itself only.

In the crucial region $r = R_{res} \sim 700$ $km$
obeying the resonance condition \eq{resonance}
the index $m$ is never 10/3, which would be necessary
for adiabatic neutrino conversions, $\kappa\gg 1$ in
\eq{adiabaticity1}. Thus, for our profiles
\eq{densityprofile} and \eq{magprofile} we need to
consider neutrino conversions in the non-adiabatic
regime. For the neutrino transition probability
describing non-adiabatic
$\bar{\nu}_e \leftrightarrow \bar{\nu}_{\mu, \tau}$
flavour conversions we use the same as for the
corresponding $\nu_e\leftrightarrow \nu_{\mu}$
MSW conversions \cite{Rosen},
\be[nonadiabat]
P_{\bar{\nu}_e \leftrightarrow \bar{\nu}_{\mu}} = \frac{1}{2}
[ 1 - (1 - 2P_{LZ})\cos 2\theta\cos 2\theta_m],
\ee
where the Landau-Zener probability
\be[LZ]
P_{LZ} = \exp \Bigl (-\frac{\pi}{4}\kappa F(\tan^2\theta)\Bigr )
\ee
in the case of the density profile \eq{densityprofile} differs
from the linear Landau-Zener potential by the correction factor
$F(\tan^2\theta)$ \cite{Kuo},
\be[extra]
F(\tan^2\theta) \simeq (1 - \tan^2\theta)\Bigl [1 + \frac{1}{5}\Bigl (
\ln (1 - \tan^2\theta) + 1 - \frac{1 + \tan^2\theta}{\tan^2\theta}\ln
(1 + \tan^2\theta)\Bigr ) +...\Bigr ].
\ee
due to the fact that in our case $V\sim \rho \sim r^{-5}$ instead of
$V_{LZ}\sim r$. We now turn to our main application of \eq{nonadiabat}.
It has been recently argued \cite{SSB} that the non-observation
of a hard energy tail in the electron anti-neutrino spectrum from
SN1987A in the Underground Detectors at the Kamiokande and IMB experiments
may place stringent limits in neutrino oscillation parameters.
In ref. \cite{SSB} this argument was used in order to
severely constrain the possibility of large mixing
MSW solutions to the solar neutrino problem.
We now proceed to the implications of their argument
to our case. Using \eq{adiabaticity1} to \eq{extra} and adapting
the results of ref. \cite{SSB} to our resonant
$\bar{\nu}_e\leftrightarrow \bar{\nu}_{\mu}$ flavour
conversions we obtain the following constraint
on the active-active light neutrino mixing parameters,
\be[bound]
\exp \Bigl (-\frac{\pi}{4}\kappa F(\tan^2\theta)\Bigr ) \gsim \frac{0.3}{2
\cos 2\theta \mid \cos 2\theta_m \mid } + 0.5,
\ee
which is shown in Fig. 1. For the case of small mixing,
$\theta \ll 1$, this bound reduces to
$\exp (- \pi\kappa /4)\gsim 0.65$. We see from Fig. 1
that assuming strong random magnetic field generation
so as to realize our new resonant anti-neutrino conversion
mechanism in SN1987A we can exclude here the possibility
of observing neutrino oscillations in all previous
searches at accelerators as well as reactors.

It is instructive to compare our results with
those of ref. \cite{SSB}. In our case one can write
\be[wolf]
\sin^2 2 \theta_m = \frac{\tan^2 2 \theta}
{\tan^22\theta + [ 1+( V + 2 \mu_{eff} B_\parallel ) / (\Delta \cos
2\theta)]^2}
\ee
as the generalization of Wolfenstein's formula
describing anti-neutrino mixing in a magnetized
degenerate electron gas. Clearly, outside the
resonance region $\sin^2 2 \theta_m \ll 1$
for neutrino masses in the range of interest
for the explanation of the solar neutrino
deficit. In this case we can safely neglect
anti-neutrino conversions outside the
resonance region. The situation in quite opposite
in ref. \cite{SSB}, where the anti-neutrino
conversion happen in vacuo outside the
star. As a result Bahcall, Spergel and Smirnov can
exclude only large vacuum mixings, in contrast to
our formula \ref{bound}.

%%%%%%%%%%%%%%%%%%%%%%%%%%%%%%%%%%%%%%%%%%%%
\section{Discussion and conclusions}
%%%%%%%%%%%%%%%%%%%%%%%%%%%%%%%%%%%%%%%%%%%%

We have found new resonant mechanism of neutrino
conversion induced by the presence of magnetisation
effects caused by axial vector interactions of
neutrinos with the charged leptons in the
degenerate electron gas in a large-scale
supernova magnetic field. We gave an explicit
solution of the corresponding evolution equation
for the case of two neutrino species. For the
conversion probalities describing
$\bar{\nu}_e \leftrightarrow \bar{\nu}_{\mu, \tau}$
flavour conversions,  we use the same results
obtained in ref. \cite{Rosen} as for the
corresponding $\nu_e\leftrightarrow \nu_{\mu}$
MSW conversions.

Our new resonance may affect both
anti-neutrino as well as neutrino flavour conversions,
and also those involving sterile neutrinos $\nu_s$.
They may occur even in situations where the MSW effect
can not occur. In particular, they may convert supernovae
anti-neutrinos $\bar{\nu_e} \leftrightarrow \bar{\nu}_b$
where $b = \mu, \tau$ at the same time that solar
neutrinos are converted through the usual
${\nu_e} \leftrightarrow {\nu}_b$ MSW conversions
and for the same choice of parameters.

Supernova neutrino energy spectra
may be substantially affected by our new resonance.
Using SN1987A data we conclude that only laboratory
experiments with long baseline such as ICARUS or MINOS
or the new generation of reactor long baseline
experiments Chooz and San Onofre are likely to
observe neutrino oscillations due to their
sensitivity to small $\Delta m^2$. Our result
is totally complementary to the one found by
Bahcall, Smirnov and Spergel. In their case
the solar neutrino conversions occur in the resonant
regime but the supernova antineutrino ones
do not. As a result they are only able to
exclude large mixing angles but with a better
sensitivity to small $\Delta m^2$. On the other
hand, if our new resonance takes place, we can
have both solar neutrinos and supernova anti-neutrinos
resonantly converted, the first by the MSW effect and the
second by our new effect involving the magnetisation.
Correspondingly we can exclude a much larger region,
including small and intermediate mixings.

Finally, let us comment on the possibility of resonant
conversions induced by Majorana neutrino transition moments.
This case is analogous to the generalisation of the
Aneziris-Schechter twisting magnetic field
($B_{\perp} = B_{0\perp} \exp (i \Phi)$) result
\cite{AnezirisSchechter} by Smirnov \cite{Smirnov},
who included matter effects. For definiteness let us
focus here on transitions to a sterile neutrino $\nu_s$.
The frequency for $\nu_e \leftrightarrow \nu_s$ transitions
\be[frequencynew]
\omega_0 = \sqrt{(V - \Delta +  \mu_{eff}B_{\parallel})^2
+ (2\mu B_{\perp})^2}
\ee
contains the magnetisation term $\mu_{eff}B_{\parallel}$
instead of $\dot{\Phi}$. The corresponding resonance condition is
\be[res]
V - \Delta + \mu_{eff}B \cos\alpha = 0.
\ee
Notice that, in contrast to ref. \cite{Smirnov},
in order to obtain the resonance condition \eq{res},
all we need is a restriction on the value of
magnetic field $B$, which enters through the term
$\mu_{eff} B_{\parallel}$. Our effect would seem
more physical, as it is generated by the charged
leptonic axial interaction in the medium and does
not depend on special field geometry details.
Note also that our axial term cures the suppression
effect for $\Delta = 0$ found in ref. \cite{VVO}
for Dirac neutrino spin flip in an external magnetic
field $\nu_{eL}\leftrightarrow \nu_{eR}$.

%%%%%%%%%%%%%%%%%%%%%%%%%%%%%%%%%%%%%%%%%%%%
%%%%%%%%%%%%%%%%%%%%%%%%%%%%%%%%%%%%%%%%%%%%
\vfill
\noindent {\bf Acknowledgment}\\
\vskip 1cm

We thank John Bahcall, Peter Goldreich, Stanislav Mikheev,
Hiroshi Nunokawa, Andreas Reisenegger, Alexandre Rez, David
Schramm, Gunter Sigl,
Dmitri Sokoloff for helpful discussions and correspondence.
This work was supported by DGICYT under grant numbers PB92-0084
and SAB94-0325. V. S. also got support from RFFR under grant
N. 95-02-03724. S. S. is supported by a fellowship
from Ministerio de Educaci\'on y Ciencia.

\vskip 1truecm

\newpage
\begin{center}
\vskip 5cm
{\bf Figure Captions}
\end{center}
\vskip2cm

{\bf Fig.1.} \\
Constraints on neutrino parameters $\Delta m^2$ and
$Sin^22\theta$ for typical average supernova neutrino
energies $<E_{\nu}>$ and different magnetic field profiles.
The curves correspond to the magnetic field profile
$B=B_0(10km/r)^m$. There are three pairs of almost
identical curves, with the lower curve in each pair
corresponding to $m=1/2$ and the upper one to $m=3/2$,
illustrating the insensitivity of our conversion rates
to the field profile. The three sets of curves correspond
to $<E_{\nu}>$ = 11 MeV, 16 MeV and 25 MeV.

\newpage
\ignore{
\begin{figure}
\centerline{
\psfig{file=/users/valle/sahufig.ps,width=.8\textwidth}}
%\caption{
\end{figure}
}
\eject
%%%%%%%%%%%%%%%%%%%%%%%%%%%%%%%%%%%%%%%%%%%%

\end{document}